\documentclass[%
 reprint,
 amsmath,amssymb,
 aps,
]{revtex4-1}

\usepackage{graphicx}
\usepackage{dcolumn}
\usepackage{bm}

\begin{document}

\preprint{APS/123-QED}

\title{Homeorhesis in Waddington's Landscape by Epigenetic Feedback Regulation}

\author{Yuuki Matsushita}
\author{Kunihiko Kaneko}
 \email{kaneko@complex.c.u-tokyo.ac.jp}
 \altaffiliation[Also at ]{Research Center for Complex Systems Biology, Universal Biology Institute, University of Tokyo}
 
\affiliation{%
Department of Basic Science, Graduate School of Arts and Sciences, University of Tokyo, 3-8-1 Komaba, Meguro-ku, Tokyo 153-8902, Japan
}%

\date{\today}

\begin{abstract}
In multicellular organisms, cells differentiate into several distinct types during early development.
Determination of each cellular state, along with the ratio of each cell type, as well as the developmental course during cell differentiation are highly regulated processes that are robust to noise and environmental perturbations throughout development.
Waddington metaphorically depicted this robustness  as the epigenetic landscape in which the robustness of each cellular state is represented by each valley in the landscape.
This robustness is now conceptualized as an approach toward an attractor in a gene-expression dynamical system.
However, there is still an incomplete understanding of the origin of landscape change, which is accompanied by branching of valleys that corresponds to the differentiation process.
Recent progress in developmental biology has unveiled the molecular processes involved in epigenetic modification, which will be a key to understanding the nature of slow landscape change.
Nevertheless, the contribution of the interplay between gene expression and epigenetic modification to robust landscape changes, known as homeorhesis, remains elusive.
Here, we introduce a theoretical model that combines epigenetic modification with gene expression dynamics driven by a regulatory network.
In this model, epigenetic modification changes the feasibility of expression, i.e., the threshold for expression dynamics, and a slow positive-feedback process from expression to the threshold level is introduced.
Under such epigenetic feedback, several fixed-point attractors with distinct expression patterns are generated hierarchically shaping the epigenetic landscape with successive branching of valleys.
This theory provides a quantitative framework for explaining homeorhesis in development as postulated by Waddington, based on dynamical-system theory with slow feedback reinforcement.
\end{abstract}

\maketitle


\section{Introduction}

\begin{figure}[htb]
\centering
\includegraphics[width=0.8\linewidth]{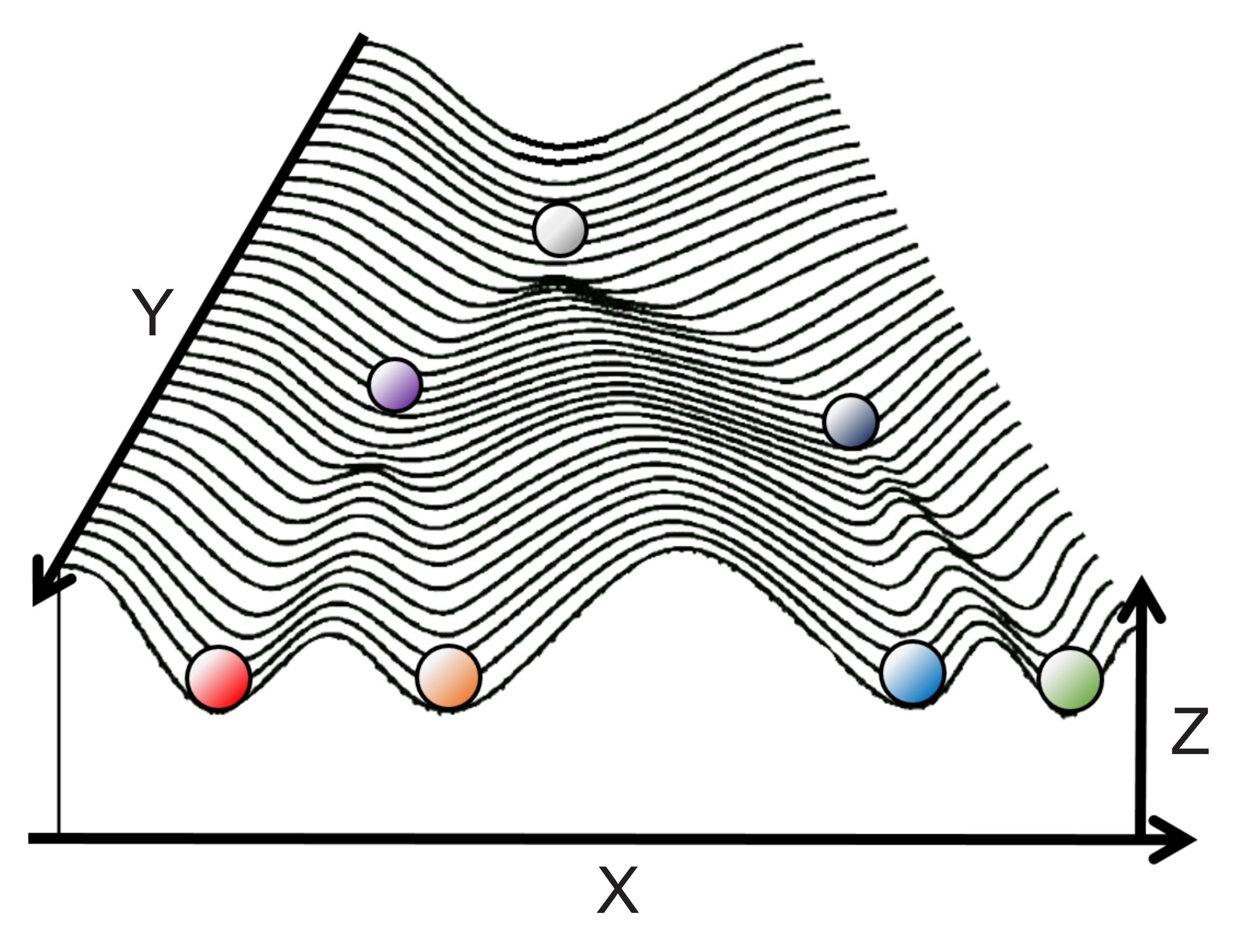}
\caption
{
Waddington's epigenetic landscape.
The cell differentiation process is conceptually explained as motion of a ball along the landscape in which valleys correspond to differentiated cell types.
Here, the horizontal axis ($X$) represents a cellular state, the height ($Z$ axis) represents the inverse of the frequency (probability) that a cell takes state $X$, and the $Y$ axis represents slow developmental change.
Adapted from \cite{waddington1957strategy}.
}
\label{Waddington}
\end{figure}

In most multicellular organisms, cells differentiate into several types in the course of development, which show distinct gene expression patterns that are robust to external perturbations and internal noise.
As a theoretical explanation for this robustness,  Waddington introduced the concept of the ”epigenetic landscape” more than 60 years ago, as shown in Fig.\ref{Waddington}. In this concept, a ball falling along the landscape represents the cell differentiation process over time, and each valley corresponds to a differentiated cell type \cite{waddington1957strategy}.
Although presented visually as a metaphor, Waddington also proposed that this differentiation process can be understood in terms of the dynamical systems of gene expression.
Following his insight, each valley is now interpreted as an attractor of an intracellular dynamical system for gene (protein) expression.
Each state remains in the vicinity of the attractor under internal noise or external perturbation.
In fact, several dynamical-systems models with mutual activation and inhibition of protein expression demonstrated the coexistence of multiple attractors that correspond to distinct cell types, and confirmatory experiments have been carried out \cite{kauffman1969metabolic, wang2010potential, wang2011quantifying, mojtahedi2016cell, furusawa2012dynamical, forgacs2005biological}.

According to this dynamical-systems approach, the $X$ axis characterizing the cellular state in Fig.\ref{Waddington} is represented  by the gene expression pattern.
However, since there are thousands of genes (or components) in a cell, the state may not be accurately represented by a one-dimensional variable $X$.
Nevertheless, the cellular state can potentially be represented by only a few variables extracted from data reduction of the expression levels of a huge number of components, such as principal component analysis (PCA) \cite{huang2005cell}.

Moreover, the height of the landscape ($Z$ axis) represents changeability of the state.
Cellular states are attracted to the bottom of the valley, which, in terms of dynamical systems, are fixed-point attractors at which point no more change will occur.

Along with the dynamics falling onto the bottom of the valley, as represented by motion along the $X$ axis in Fig.\ref{Waddington}, the landscape itself is shaped along the other ($Y$) axis representing the developmental course, in which the valleys are shaped successively and are deepened, in a process known as “canalization”.
Therefore, a fundamental question remains:  given that the attraction to each valley along the $X$ axis is represented by gene expression dynamical systems, what does the $Y$ axis representing (slower) landscape change represent?

To address this fundamental question, there are three basic questions to resolve with respect to the postulates of Waddington's landscape itself.

First, there is the issue of hierarchical branching. That is,
since the valleys are successively generated over developmental time (Fig.\ref{Waddington}),many valleys (attractors) are not generated independently, but rather the shallower valleys are generated first and are then branched, and these branching processes are repeated \cite{wagner2018single, briggs2018dynamics}.
 
Second, Waddington argued that the developmental process itself, i.e., the motion along the shaping of valleys, is also robust to perturbation, and coined the term “homeorhesis” to represent such path stability \cite{waddington1957strategy}.
However, the mechanism contributing to the robustness of this shaping process, including successive branchings, remains elusive.

Finally, the number ratio of each cell type is also rather robust to perturbations or initial conditions.
If we assume that a deeper valley attracts more cells, this robustness implies overall robustness of the landscape, in particular, the depth of each valley.

Considering these three postulates of the landscape, let us now come back to the fundamental question of the nature of the $Y$ axis representing slow developmental change.
After 60 years since proposal of the epigenetic landscape, we have now identified possible candidates that could cause such a slow landscape change.
One candidate is the cell-cell interactions \cite{kaneko1999isologous, furusawa2001theory, suzuki2011oscillatory}.
As development progresses and the cell number increases, the influence of cell-cell interactions on the intracellular dynamics for each type becomes stronger. Slow modifications of intracellular expression dynamics can lead to novel attractors or the increase in their robustness.

Another potential source of slow landscape change is epigenetic modification \cite{waddington1942epigenotype, bird2007perceptions}, which is currently one of the hottest topics in cell and developmental biology \cite{goldberg2007epigenetics, jaenisch2003epigenetic, surani2007genetic, huang2009non, cortini2016physics, singer2014dynamic}.
Epigenetic changes such as histone modification or methylation are now well established as an essential mechanism to the feasibility of the expression of each gene at a given time and place \cite{schwartz2007polycomb, gombar2014epigenetics, rohlf2012modeling, karlic2010histone}.
The epigenetic change itself is slow, and is stabilized through a positive-feedback process as demonstrated both theoretically and experimentally \cite{dodd2007theoretical, sneppen2008ultrasensitive, sasai2013time}.
Despite growing interest and extensive reports on epigenetic changes, the detailed interplay between epigenetic modification and gene expression dynamics has been rarely explored.
In contrast to the extensive body of theoretical and empirical literature on expression dynamics or epigenetic modifications, there is no experimental elucidation of the underlying molecular mechanisms nor theoretical model for the interplay proposed to date.
Therefore, a simple phenomenological model is needed to investigate how such slow epigenetic change can introduce a novel expression pattern or stabilize the existing expression patterns.  
Such a model would provide a bridge between epigenetic modification and the epigenetic landscape as Waddington conceptualized.

Note that epigenetic modification generally depends on a given cellular state, i.e., the expression levels of proteins, whereas the epigenetic modifications influence the expression levels and thus determine the state.
In general, the epigenetic process is slower than the expression dynamics.
The epigenetic modification leads to stabilization of cellular states, i.e., deepening the valleys  as schematically represented in Fig.\ref{Waddington}.
To formulate the epigenetic process in terms of dynamical systems, we here introduce an epigenetic variable for each expressed gene, represented as a threshold level of the input needed for the gene of concern to be expressed.

Using the simplest feedback process, we elucidate the possible conditions for the epigenetic landscape and its properties.
Rather than seeking detailed models extracted from realistic expression dynamics, we instead consider a minimal conceptual model that captures the interplay between the relatively faster gene expression dynamics and slower epigenetic dynamics to address how an epigenetic landscape satisfying the requisites of (1) hierarchical branching, (2) homeorhesis, and (3) robustness in the cell-number ratio is generated.
Instead of the simplicity in the model, we have simulated thousands of networks, to extract a universal mechanism and draw a general conclusion, which will hold true in a complicated system with biological reality.

\section{Model}
We consider a cell model with a gene regulatory network (GRN) and epigenetic modification.
The cell has $N$ genes and the cellular state is represented by the expression level (or concentration) $x_i$ of each gene $i$.
Here, the GRN represents the mutual control of genes via synthesized proteins.
Gene expression typically shows an on-off-type response to the input:
a gene is expressed (suppressed) when its input value is above (below) a certain threshold, whereas its expression level is saturated as the input value increases.
By normalizing the maximal (i.e., saturated) expression level to unity, we adopt the following gene expression dynamics for simplicity \cite{mjolsness1991connectionist, salazar2000gene, salazar2001phenotypic, kaneko2007evolution, Jia_2019}:
\begin{align}
\frac{d x_i}{d t} = F \left( \sum_j^N \frac{J_{ij}}{\sqrt{N}} x_j + \theta_i  + c_i \right) - x_i \label{dx},
\end{align}
where $J_{ij}$ is the regulatory matrix.
If $J_{ij}$ is positive (negative), gene $j$ activates (represses) the expression of gene $i$, whereas $J_{ij}$ is set to 0 if no regulation exists.
To represent the on-off-type expression of genes, we adopt $F(z) = \tanh (\beta z)$
\footnote
{
This form is adopted for the convenience of analysis, but using the form $F(z)=(\tanh (\beta z)+1)/2$ to set the expression level between [0,1], or adopting the Hill function such as  $F(z) = z_i^{\alpha}/(z_i^{\alpha} + \theta_i^{\alpha})$ does not qualitatively alter the result.
}
($\beta = 40$).
$x_i = 1$ indicates the full expression of the $i$ th gene and $x_i = -1$ indicates no expression of the $i$ th gene.
$c_i$ is a constant input value, interpreted as an input outside of the $N$ genes (for example, upstream genes) or the natural trend for expression.
For most examples, however, $c_i$ is set to 0 unless otherwise noted.

Here, $-\theta_i$ represents the threshold for the input, beyond which the expression is activated.
As $\theta_i$ is increased (decreased), the $i$-th gene tends to be expressed (repressed). 
In the standard GRN model, this threshold is fixed. By contrast, we regard it as a variable by assuming that $\theta_i$ represents the epigenetic modification level for each gene $i$, such as histone modification or DNA methylation. 
Further, the epigenetic modification is changed depending on the expression level of gene $i$.
In accordance with some experimental reports \cite{spainhour2019correlation, dodd2007theoretical, sneppen2008ultrasensitive}, we adopt a positive-feedback process from gene expression to epigenetic modification; that is,
when a gene is expressed (repressed), it becomes easier (harder) to be expressed, as given by:
\begin{align}
\frac{ d \theta_i}{d t} = v \left( a x_i - \theta_i \right). \label{dtheta}
\end{align}
In \eqref{dtheta}, the parameter $a( > 0)$ represents the strength of the positive-feedback mechanism, and $v ( > 0)$ gives the rate of change in the epigenetic modification.

\section{Fixed-point analysis}
The fixed-point solutions of \eqref{dx} and \eqref{dtheta} are obtained by setting each term to zero.
From the latter, we get $\theta_i = a x_i$ and from the former we obtain
\begin{align}
\tanh \beta \left( \sum_i^N \frac{J_{ij}}{\sqrt{N}} x_j^* + a x_i^* \right) - x_i^* = 0 \label{fixed_point}
\end{align}
(note that the case with $c_i = 0$ is considered here).
In the large $\beta$ limit, the $\tanh$ function is approximated by the step function, so that the fixed point $x_i^*$ is given by a sequence of $\{-1, 1\}$ that satisfies \eqref{fixed_point}.
The number of fixed points of \eqref{fixed_point} then increases monotonically with the value of $a$ (Fig. S1).
If it is large enough (that is, the second term in the brackets in  \eqref{fixed_point} is sufficiently larger than the first term), all of the $2^N$ patterns with any combination of $x_i^* = \pm 1$ (with $\theta_i^* = ax_i^*$) satisfy \eqref{fixed_point}.
All of these are fixed-point attractors, which are reached by choosing initial conditions close to each $\{ -1, 1\}^N$ state.
However, for $a = 0$, the number of fixed points satisfying  $x_i^* = \tanh \beta (\sum J_{ij} x_j^*/\sqrt{N})$ is much smaller.

Here, we focus on a case with sufficiently strong epigenetic feedback, i.e., sufficiently large $a$, in which all of the possible $2^N$ states could exist if any value of $x_i$ and $\theta_i$ is initially chosen.
However, for studying the canalization dynamics, we restrict the initial condition of $\theta_i$ as follows: At the initial stages of development, epigenetic modification is not yet introduced \cite{feldman2006g9a, challen2012dnmt3a, reik2001epigenetic, hawkins2010distinct, PhysRevX.9.041020}, so that all of $\theta_i$s are set to $0$.
Under this restriction, we investigate which of the $2^N$ fixed points with $x_i^* = \pm 1$ and  $\theta_i^* = ax_i^*$ is reachable through developmental change of the epigenetic modification.
As we limit the dynamics to the state of $\theta_i = 0$, we refer to only the final states reached from such initial conditions as attractors throughout the paper (whereas the initial conditions of $x_i$s cover all possible $\{-1, 1\}$ states).

\section{Attractor generation and pruning}
First, we set $N = 10$ and prepare the initial conditions for all gene expression patterns with null epigenetic modification (i.e., $2^N$ candidates with $x_i = \pm1, \theta_i = 0$).
In the context of the epigenetic landscape, these initial conditions correspond to the balls on the top of the landscape, whereas the valleys are shaped with the change in $\theta_i$ and the balls are trapped at the bottom of the landscape that correspond to the attractor.
We then examine which and how many attractors are selected depending on the parameter $v$.

At the limit of $v \rightarrow 0$, i.e., the adiabatic limit in terms of physics, the time scales of the dynamics for $x_i$ and $\theta_i$ are well separated.
Only after the expression level $x_i$ reaches one of the original attractors with $\theta_i$ originally fixed at $0$, $\theta_i$ begins to show gradual variation.
Hence, the number of attractors will be bounded by the expression dynamics when fixing $\theta_i = 0$.
At the limit of $v \rightarrow \infty$, $\theta_i$ reaches $\theta_i = ax_i$ faster, so that all of the $2^N$ states $x_i^* = \tanh (\beta a x_i^*)$ are attracted depending on the initial $x_i$ values, as long as $a$ is sufficiently large.
By considering these two extreme limits, $v$ generally functions as a parameter that limits the final state from all of the possible $2^N$ states.
Now, from naive expectation based on the above two limits, it might be expected that the number of attractors will monotonically increase with $v$.
Indeed, such monotonic increase could be observed for 80\% of randomly chosen networks $J_{ij}$ for $N = 10$.

For $v\sim 0$, the approach to the attractor is completed before epigenetic modification and then $\theta_i$ is fixed accordingly. 
With the introduction of $v$, $\theta_i$ increases or decreases depending on the initial value of $x_i$. If this process for $x_i$ is fast, $x_i$ is fixed to $\pm 1$ depending on the initial condition; that is, before the approach to the original attractor.
Hence, the original basin of attraction is partitioned. With the increase in $v$, more partitions progress; accordingly, the few attractors that exist at $v=0$ are successively partitioned toward $2^N$ states with the increase in $v$.
In this case, for a given $v$, fixation simply occurs from the neighborhood of each on/off-pattern attractor provided by the initial condition.
There exists no hierarchical branching to each attractor over developmental time.
Moreover, since only the attractor from the neighborhood of the initial expression state is reached, the final state crucially depends on the initial condition, the final state crucially depends on the initial condition.
Neither homeorhesis nor robustness in the cell-number ratio is expected, as will be confirmed later.

However, in the case of $N = 10$, approximately 20 \% of the randomly chosen matrix $J_{ij}$ shows non-monotonic dependency of the attractor number on $v$. Here, different attractors are generated and pruned successively with $v$ in the intermediate range of $v$. This implies that states separated at smaller $v$ converge again with the increase in $v$, even though the epigenetic feedback tend to separate each $x_i$ to $\pm 1$. With mutual interference between the fast dynamics of $x_i$ and slower dynamics of $\theta_i$, both the convergence of initial states and divergence to fixed states coexist, as will be discussed below.
Further, as will be shown, such convergence of orbits in the initial regime can allow for creation of an epigenetic landscape that satisfies the three postulates of hierarchical branching, homeorhesis, and robustness in the cell-number ratio.

\begin{figure}[htb]
\centering
\includegraphics[width=\linewidth]{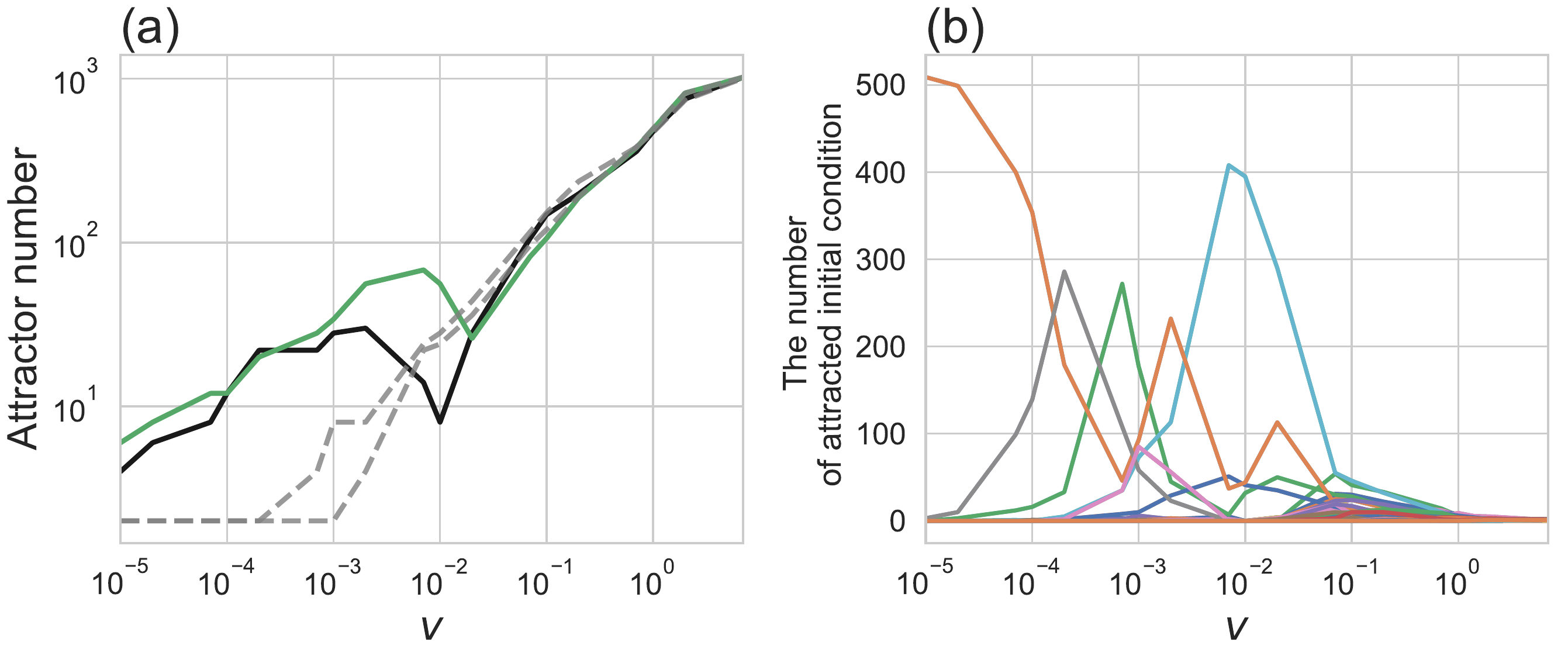} \\
\caption{
(a) Dependence of the number of attractors (reached states from $\theta_i = 0$) upon $v$.
Grey dotted lines show the case with a monotonic increase of the attractor number against $v$.
The black and green solid lines are examples with non-monotonic dependence on  $v$.
Attractors are pruned at $2\times 10^{-3} < v < 1\times 10^{-2}$.
$N=10$.
(b) Dependence of the basin volume of each attractor upon $v$, for the example of non-monotonic dependence of attractor number shown as the black line in (a).
Basin volume is computed by taking $2^{N}$ initial conditions of $\{x_i=\pm 1 \}$ and setting  $\theta_i=0$ initially, and then counting the number of initial conditions reaching each fixed-point state.
Each line with a different color shows the basin volume for each different attractor.}
\label{result}
\end{figure}

In this non-monotonic case, the basin volume of each attractor, i.e., the fraction of initial conditions from which each attractor is reached, also changes with $v$.
In particular, dominant attractors successively change with $v$ as shown in Fig.\ref{result}b.
This scenario is in stark contrast with the case of a monotonic increase in attractor number, where each basin of attraction is simply partitioned to $2^N$ successively with the increase in $v$ (Fig.S2).

\section{Trajectory separation by epigenetic modification: simplest example}
\begin{figure*}[htb]
\centering
\includegraphics[width=0.9\linewidth]{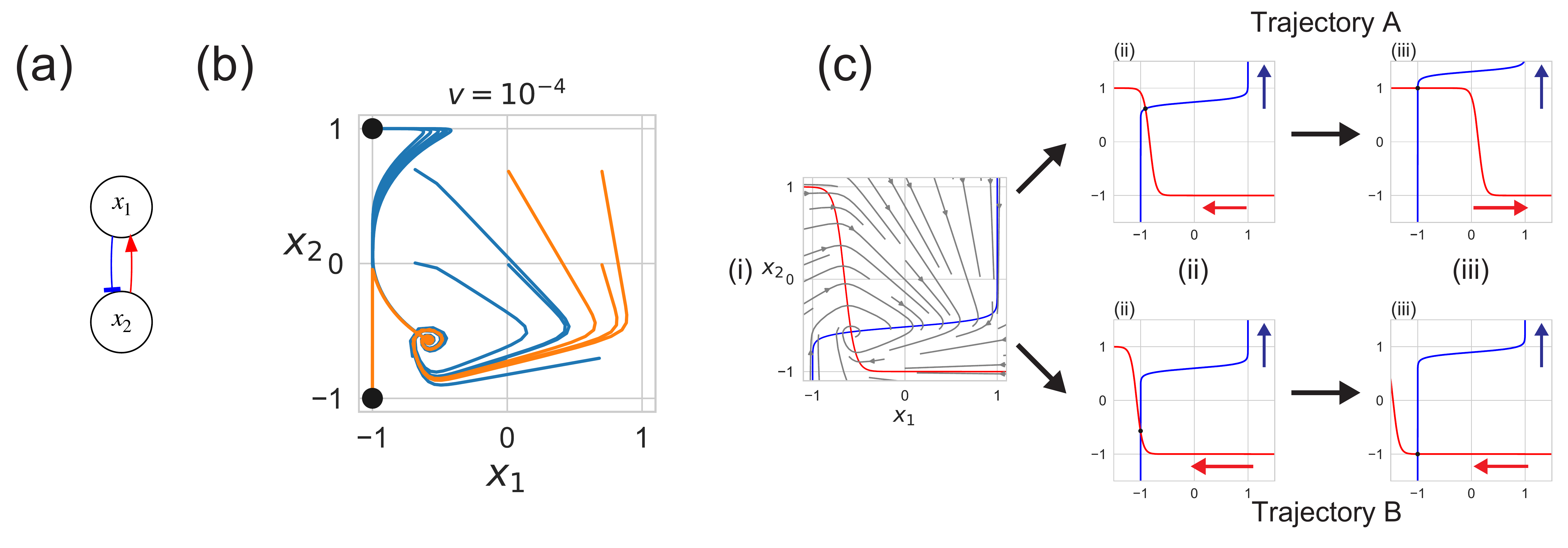}
\caption{
(a) Two-gene system with activation (red arrow) and inhibition (blue arrow):
$x_2$ activates $x_1$ and $x_1$ inhibits $x_2$.
The parameter values are chosen as $J_{12} = 0.44, J_{21} = -0.33, J_{11} = J_{22} = 0, c_1 = 0.16, c_2 = -0.15$.
(b) Trajectories in $x_1,x_2$ for $v = 10^{-4}$. Trajectory A reaches the fixed point $(-1, 1)$, whereas Trajectory B reaches $(-1, -1)$.
These two types of trajectories coexist (two fixed points as two black dots), depending on the initial condition, for the intermediate value of $v$.
Initial conditions are chosen at even intervals per 0.5 in the phase space of $(x_1, x_2)$.
(c) Analyses of the two types of trajectories according to the motion of two nullclines: blue, corresponding to $dx_1/dt=0$; red, $dx_2/dt=0$.
(i) $(x_1, x_2)$ approaches the crosspoint of the nullclines if $\theta_i$ is fixed, whereas the change in $\theta_i$ results in a shift of the nullclines.
(ii) For both trajectories, $x_1$-nullcline (blue line) goes up and $x_2$ nullcline (red line) goes left first, because $(x_1,x_2)$ first approaches the fixed point at $\theta_i = 0$ starting from any initial condition. Upper: As $x_2$ exceeds $0$, the motion of the $x_1$ nullcline changes its direction, and $(x_1, x_2)$ reaches the fixed point $(-1, 1)$.
This gives Trajectory A. Lower: Before $x_2$ reaches 0, the $x_2$ nullcline crosses the $x_1$ nullcline vertically in Trajectory B so that $x_2$ remains negative and the motion of the nullclines do not change their direction; thus, $(x_1, x_2)$ reaches $(-1, -1)$.
}
\label{Two-gene system}
\end{figure*}

To understand how mutual feedback between gene expression and epigenetic modification can lead to the generation and pruning of attractors, we first consider the minimal case with only two genes ($N = 2$).
In addition, $c_1, c_2 \ne 0$, which may be also regarded as the inputs from genes other than $i=1,2$.
We consider the case $J_{11} = J_{22} = 0, J_{12} > 0 > J_{21}$; i.e., one gene activates the other, which then inhibits the first, as shown in Fig.\ref{Two-gene system}a.

In this simple case, the number of attractors changes as $1 \rightarrow 2 \rightarrow 1$ with the increase in $v$ over a certain range of parameters $c_1, c_2$ (Fig. S3a).
For $v < 5.7\times10^{-5}$, only trajectories reaching $(-1, 1)$ are realized (Trajectory A)(Fig. S3b).
By increasing $v$ further, trajectories reaching $(-1, -1)$ then appear (Trajectory B) where $5.7\times10^{-5} < v < 7.0\times10^{-4}$,  and two attractors $(-1, 1), (-1, -1)$ coexist (Fig.\ref{Two-gene system}b).
For larger $v$ ($7.0\times10^{-4} < v < 9.1\times10^{-3}$), the attractor $(-1, 1)$ disappears completely (Fig. S3c).
The time course in the development of the two types of trajectories and the change in the basin for each attractor are shown in Fig. S4 and Fig. S5, respectively.

The above $v$ dependency of attractors is explained as follows.
When $v$ is small, the dynamics are approximated by the means of "adiabatic elimination"; i.e., $x_i$ reaches the fixed point for a given $\theta_i$, whereas $\theta_i$ changes slowly.
For given $\theta_i$, the $\{x_i\}$ dynamics are analyzed by the two nullclines, given by
\begin{align}
dx_1/dt = 0 \ \rightarrow \ x_1 = \tanh \beta \left( \frac{J_{12}}{\sqrt{N}} x_2 + \theta_1 + c_1 \right),\\
dx_2/dt = 0 \ \rightarrow \ x_2 = \tanh \beta \left( \frac{J_{21}}{\sqrt{N}} x_1 + \theta_2 + c_2 \right).
\end{align}
When $v$ is small, while $x_i$ moves towards the crosspoint of the two nullclines, as $\theta_i$ slowly changes according to \eqref{dtheta}, the nullclines are slowly shifted.

When this adiabaticity condition is satisfied, only Trajectory A is realized (Fig. S3b): 
at $\theta_i = 0$ (null epigenetic modification), there is a stable fixed point as the crosspoint of the two nullclines at $x_1 < 0$ and $x_2 < 0$ (Fig.\ref{Two-gene system}c(i)).
Then, according to \eqref{dtheta}, each nullcline is shifted as follows: 
the $x_1$-nullcline (i.e., $d x_1/d t = 0$ nullcline) goes up, whereas the $x_2$-nullcline (i.e., $d x_2/d t = 0$ nullcline) goes left.
As a result, the crosspoint of the two nullclines itself moves up and left, thus reaching above $x_2 = 0$.
Consequently, the shift of the $x_2$-nullcline changes its direction (as the sign of $d \theta_2/d t$ is approximately given by the sign of $x_2$).
Accordingly, the crosspoint of the nullclines continues to move up, reaches $(-1, 1)$, and then stops.

However, by increasing $v$, the faster move of the nullclines generates another trajectory, Trajectory B.
First, the crosspoint of the two nullclines moves to the left and up, in the same way as observed for Trajectory A.
However, owing to the faster change in $\theta$, the $x_2$-nullcline shifts to the left so quickly that the two nullclines cross vertically (see Fig.\ref{Two-gene system}c (ii) Trajectory B), and the crosspoint does not go above $x_2 = 0$.
As a result, the crosspoint moves to the left and down to $(-1, -1)$, where $(x_1, x_2)$ is fixed for some initial conditions.
Here, $(x_1, x_2)$ first approaches the fixed point at $\theta_i = 0$ for both Trajectories A and B, and then owing to slight difference in the initial conditions, $(x_1, x_2)$ is directed either to $(-1, 1)$ or $(-1, -1)$.

By increasing $v$ beyond $9.1\times10^{-3}$, the shift in the nullclines is accelerated, so that the two nullclines cross vertically for all of the initial conditions.
In this case, Trajectory A is not realized for any initial condition, and all of the initial conditions are instead attracted to $(-1, -1)$ (Fig. S3c).

Hence, the attractor number increases due to the divergence in the motion of the nullclines depending on the initial conditions of $\{x(i)\}$.
With a further increase in $v$, the attractor is pruned because nullclines move faster and no longer split into two directions of motions due to the faster change of $\theta_i$.

\section{Generation and pruning of attractors from an oscillatory state}
\begin{figure}[htb]
\centering
\includegraphics[width=\linewidth]{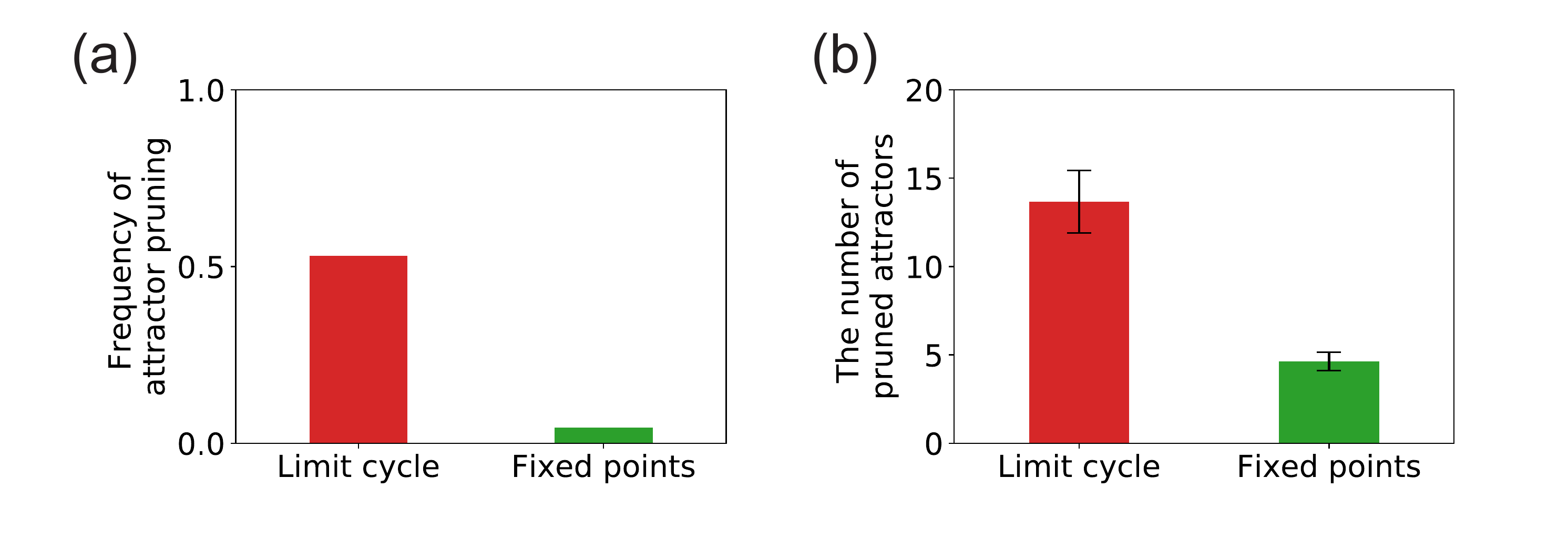}
\caption{
(a) Fraction of the regulatory matrices $J_{ij}$ that exhibit attractor pruning with the increase in $v$ (here defined as a decrease in attractor number of more than 4).
The cases with the initial limit-cycle attractors at $\theta_i=0$ (red) and those with fixed points (green) are sampled separately.
(b) Average number of pruned attractors, defined as the difference between the local maximum and local minimum of attractor number against the change in $v$.
Case with a limit cycle (red) and fixed points (green). See Fig. S6 for more details.}
\label{comparison}
\end{figure}

The two-gene minimal model described above suggests how the interplay between fast $x$ dynamics and a slow nullcline shift leads to divergence in trajectories, thereby resulting in non-monotonic change in the attractor number.
By contrast, for $N = 10$, the non-monotonic behavior of attractor number against $v$ mostly adopts a limit-cycle attractor at $\theta_i = 0$.
The frequency of networks showing such behavior is much larger for the limit-cycle case, along with the number of generated and pruned attractors in the intermediate range of $v$ (See Fig.\ref{comparison}).

This relevance of the limit cycle to the generation and pruning of multiple attractors is explained as follows.
First, as the limit cycle travels over a larger portion in the phase space of $\{x_i\}$, the variation in the change in $\{\theta_i\}$ is enhanced so that more attractors can be generated with the increase in $v$.
These attractors are generated hierarchically by branching trajectories successively, stemming from the original limit-cycle orbit.
However, with the increase in $v$, the initial oscillation is destroyed due to the faster change in $\theta$ (shift of nullclines), so that the top of the hierarchy in branching trajectories is destroyed, leading to a drastic decrease in the attractor number.

\begin{figure}[htb]
\centering
\includegraphics[width=\linewidth]{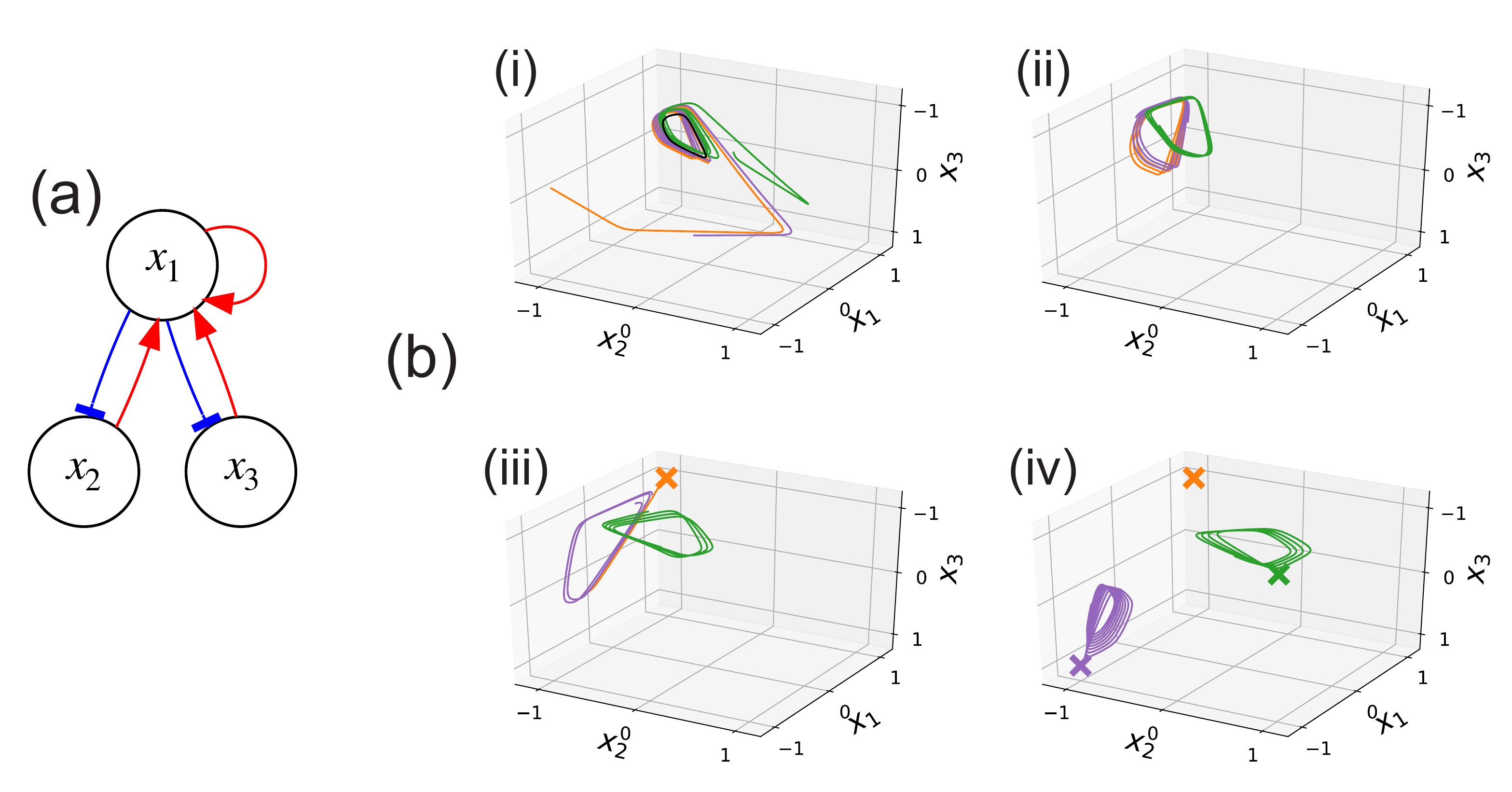}
\caption{
(a) Three-gene system with $J_{11} = 0.26, J_{12} = J_{13} = 0.35, J_{21} = -0.4, J_{31} = -0.36, c_1 = 0.26, c_2 = c_3 = 0.17$. (b) Hierarchical attractor generation from limit cycle (HAGL). 
$v = 10^{-3}$.
(i) Trajectories from different initial conditions, plotted by different colors, approach a limit-cycle attractor at $\theta_i = 0$.
(ii) Trajectories are separated into two groups (green line shows one group, and orange and purple lines show another group), depending on the initial initial condition.
(iii) Further separation of the group of trajectories shown in orange and purple.
(iv) Three trajectories reaching distinct fixed points.
}
\label{Three-gene system}
\end{figure}

This hierarchical attractor generation from limit-cycle (HAGL) is illustrated in a simple three-gene system with a limit-cycle attractor (Fig.\ref{Three-gene system}a).
In this three-gene system, only one attractor is reached for small $v$ where the adiabatic condition is satisfied (Fig. 6a).
With a further increase in $v$, however, three attractors are reached ($4\times 10^{-4} < v < 9 \times 10^{-3}$).
The trajectories reaching these attractors initially show oscillation around the original limit cycle at $\theta_i = 0$, and then separate into two groups, as shown in Fig.\ref{Three-gene system}: two fixed-point attractors are generated from one group, whereas one fixed-point attractor is generated from the other group.
Thus, the attractors are generated hierarchically.
With the increase in $v$, the initial limit-cycle orbit is destroyed before the separation into two groups, so that the number of attractors is reduced from three to one ($v \sim 9 \times 10^{-3}$) (see Fig. S7 for more details).

Most of the generation and pruning of multiple attractors can be understood as HAGL.
Note that for much larger $N$, limit-cycle attractors (or sometimes chaotic attractors) exist more often in the model \eqref{dx} with $\theta_i = 0$, as previously investigated in neural network models \cite{sompolinsky1988chaos, aoki2013slow}.
Therefore, the  generation and pruning of multiple attractors are expected to be ubiquitous.
For confirmation, we simulated the model with $N=100$.
Although sampling all $2^N$ initial conditions $\{x_i = \pm 1; i = 1, \dots , N\}$ is numerically difficult,  simulations with partial sampling showed that non-monotonic change in the attractor number occurred for most of the randomly chosen $J_{ij}$ matrices (Fig. S8) where HAGL is commonly observed.

\section{Epigenetic landscape and homeorhesis}
Thus, HAGL satisfies the first postulate of Waddington's landscape: hierarchical branching.
Now, we consider the other two postulates of homeorhesis and robustness in the cell-number ratio.
For this purpose, we first need to determine the axes $X$ and $Z$ in the landscape.

As discussed above, the $X$ axis represents the cellular state, which
can be extracted from $\{ x_i \}$ using PCA.
Here, we adopt the 1st PCA mode of $\{ x_i \}$ as $X$.
Each valley corresponds to an attractor stabilized by the slow epigenetic change.
To explore robustness in the developmental course and generated epigenetic landscape, we introduce noise in \eqref{dx} and \eqref{dtheta}. We adopt the Langevin equation by adding Gaussian white noise $\eta_i(t)$ with $<\eta_i(t) \eta_j(t')> = \sigma \delta_{ij}\delta(t - t')$, with $\delta_{ij}$ as Kronecker delta and $\delta(x)$ as a delta function.
In general, the specific attractor that is reached depends on the initial condition and perturbation by internal noise.
By taking the number of cells under noise, each cell reaches one of the attractors (and stays in its vicinity even under noise).
Then, one can compute the number distribution of $P(X)$.
As $Z$ is lower, the state with X is more frequently reached.
By analogy with the relationship between free energy and probability in thermodynamics, one can adopt $Z = \log(1/P(X))$.
Then, the epigenetic landscape can be depicted using the height $Z$ as a function of $X$.

To compute $P(X)$, we first choose an initial condition of cells (or distribution around a given initial pattern of $X_i$).
For each initial value, $X$ is computed as a result of time evolution.
By starting with a sufficient number of cells, the distribution $P(X)$ is obtained, which may depend on the initial condition of cells.
Then, to examine the robustness of the landscape, we explore whether the time evolution of the distribution $P(X)$ is robust against the change in the initial condition of cells.

First, when $v$ is large, any of the $2^N$ states is approached from the vicinity of each of the initial expression pattern $\{x_i=\pm 1 \}$.
In this case, the specific state that is attracted as well as the number distribution of cells for each state crucially depend on the choice of initial conditions.
Hence, $P(X)$ is not robust to the change in the initial conditions.

Next, we consider the case with monotonic dependence of attractor number upon $v$.
In this case, if $v$ is not so large, the number of attractors $n_A$ is much smaller.
Nevertheless, the specific attractor the cell state reaches is still predetermined by how close the expression state at $\theta_i = 0$ is to the final expression state.
The initial $x_i$ state is partitioned into $n_A$ basins, from each of which only one attractor (valley) is generated.
Hence, $P(X)$ crucially depends on the initial distribution of $x_i$'s (see Fig. S9).

In contrast, for HAGL, the postulated robustness is achieved if $v$ is in the intermediate region in which multiple attractors are generated, as shown in Fig. S9.
The obtained $P(X)$ is almost completely independent of the initial conditions of cells.
For most initial conditions, all of the attractors are reached, and the fraction of cells reaching each attractor under noise is quite stable against the change in the initial distribution of $\{ x_i \}$.
In this case, from any initial conditions, the limit-cycle attractor (at $\theta_i = 0$) is first reached. With the epigenetic feedback, the cells are then distributed to each attractor depending on the phase of oscillation.
Hence, the time course of differentiation to each attractor (cell type), as well as the fraction of each attractor (the number ratio of each cell type) are both independent of the initial distribution of cells.
\begin{figure}[htb]
\centering
\includegraphics[width=\linewidth]{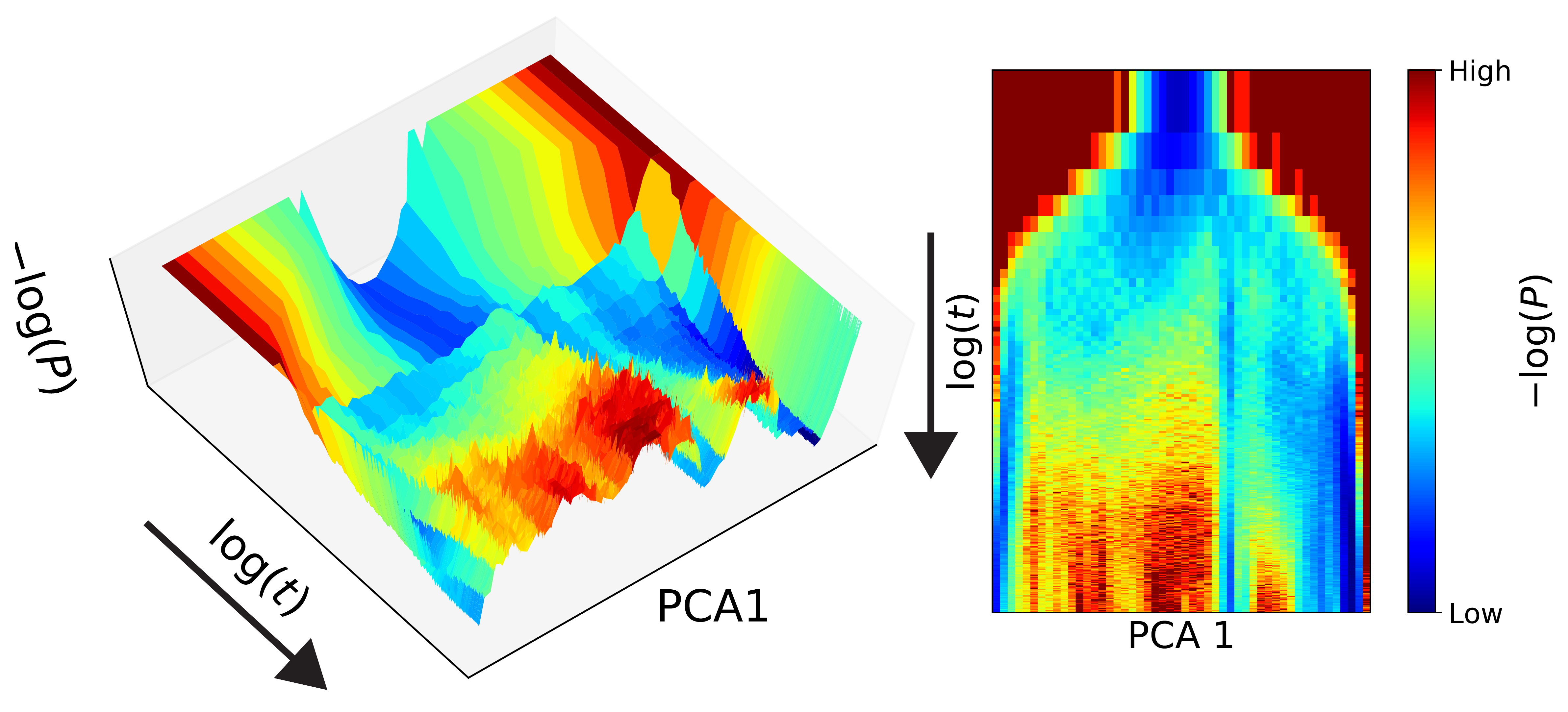}
\caption{
Epigenetic landscape generated from the temporal evolution of cellular states for given $J_{ij}$ that exhibits HAGL.
$N = 40$ and $c_i$ is set to a random value sampled from the normal distribution with average $0$, variance $0.1$.
We adopt 1-mode PCA to represent a one-dimensional scalar variable $X$ and $Z = - \log (P(X))$ indicating the depth of valley, where $P(X)$ is the distribution of $X$ over cells developed under noise, and is plotted against time given by $\log(t)$.
Red indicates large (i.e., low frequency) and blue indicates small values.
The amplitude of Gaussian white noise $\sigma = 0.1$.
The right figure shows a one-dimensional representation with the horizontal axis as $X$ and vertical axis as scaled time (from top to bottom), whereas the color represent $Z$.
}
\label{landscape}
\end{figure}

We can then depict the epigenetic landscape according to the time evolution of $P(X)$.
Here, $X$ (in Fig.\ref{Waddington}) is given by the 1st PCA mode from $\{x_i\}$ obtained from a distribution of initial conditions.
The landscape is depicted by $Z = -\log P(X)$, so that the bottom of the lower valley has a higher population density.
The landscape thus depicted is given in Fig.\ref{landscape}, which shows both the hierarchical branching and robustness to the initial expression or noise.

\begin{figure}[htb]
\centering
\includegraphics[width=7cm]{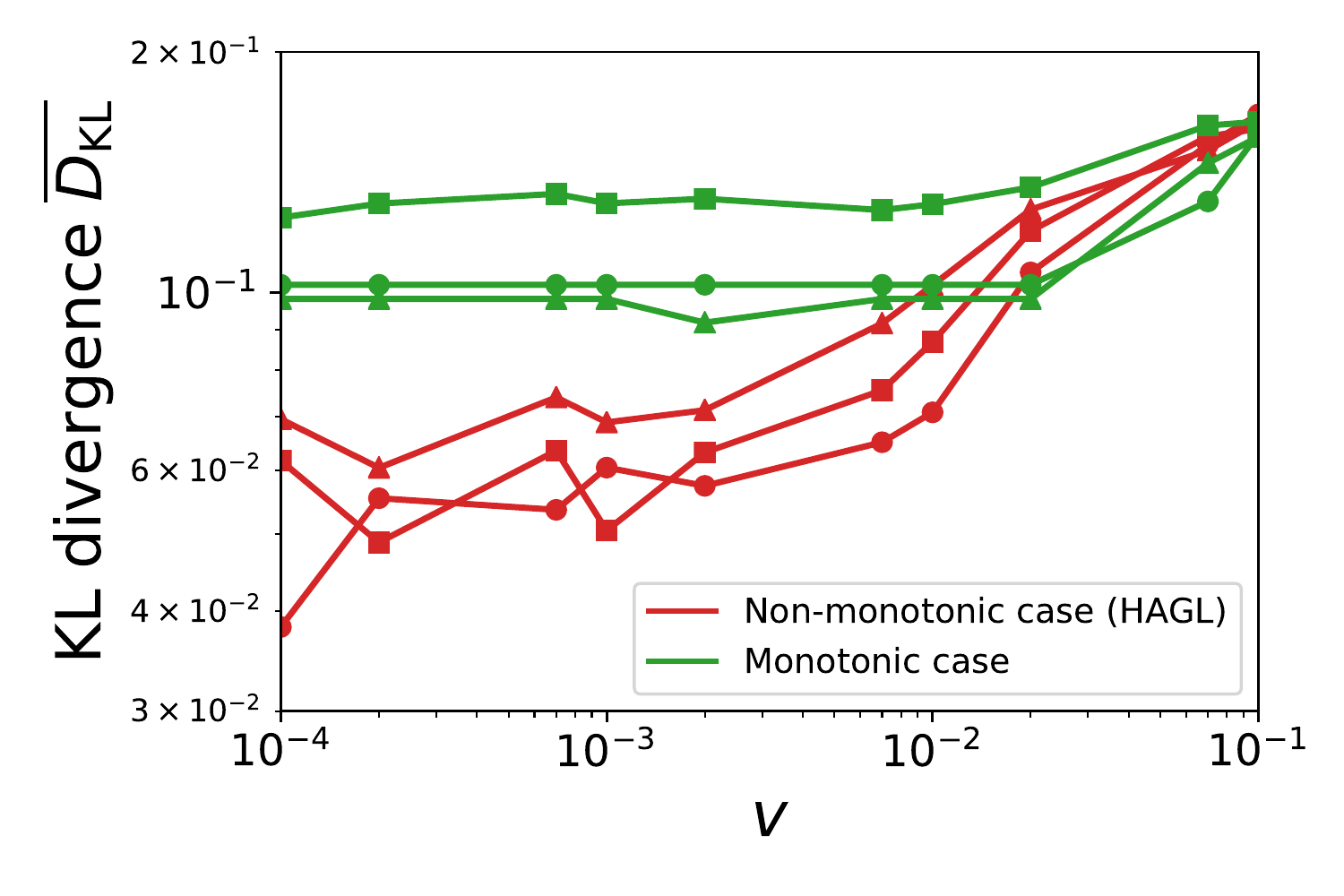}
\caption{
$\overline {D_{\mathrm{KL}}}$ representing the averaged Kullback-Leibler divergence between two distributions of cellular states developed under noise ($\sigma = 0.1$).
First, $P(X)$ is computed from 500 cells developed from a given initial condition and $J_{ij}$.
The distribution $P^{\mu}(X)$ is computed over $\mu=1, 2, \dots, 10$  starting from different initial conditions.
The Kullback-Leibler divergences are then computed over all pairs of 90 distributions and averaged to get $\overline {D_{\mathrm{KL}}}$ (see also Fig. S10 for each distribution form).
For HAGL, $\overline {D_{\mathrm{KL}}}$ remains low up to large $v \sim 10^{-3}$ (red lines), whereas in the monotonic case (green lines), it takes on a large value over the full range of $v$.}
\label{KL}
\end{figure}

Finally, we quantitatively characterize the robustness of the final distributions of cellular states reached from different initial distributions.
Let us define $P^{\mu}(X)$ as the distribution of $X$ reached from a given initial condition of $x_i$, indexed by $\mu$ (e.g., $x_i(t = 0) = \eta^{\mu}_i$, where $\eta^{\mu}_i (i = 1, \dots, N)$ is one random sequence in $[-1, 1]$, whereas $\nu \ne \mu$ denotes a different random sequence).
As the measure for the distance between two distributions $P^{\mu}(X), P^{\nu}(X)$ generated from different initial distributions, we adopt the KL divergence $D_{\mathrm{KL}} = \sum _X P^{\mu}(X) \ln \{P^{\mu}(X)/P^{\nu}(X)\}$ for a pair of two distributions $P^{\mu}(X), P^{\nu}(X)$ obtained from two samples $\mu$ and $\nu$ starting from different initial conditions.
If $D_{\mathrm{KL}}$ is small, a similar distribution $P(X)$ (i.e., a similar landscape) is obtained, independent of the initial condition, thereby implying robustness at the distribution level.
$\overline D_{\mathrm{KL}}$ is computed by averaging over the samples $\mu$ and $\nu$, which is plotted in Fig.\ref{KL} for the case of monotonic attractor number dependency on $v$ and the non-monotonic HAGL case.
As shown in Fig.\ref{KL}, the $\overline D_{\mathrm{KL}}$ value is kept small up to a large value in $v$ (e.g., $v \leq 10^{-2}$) for the HAGL case.
This quantitatively demonstrates that differentiation from the oscillatory state through epigenetic fixation shows higher robustness in the distribution of cellular states.

\section{Discussion}
We have introduced a model involving mutual interactions between the expression dynamics controlled by a GRN and epigenetic modification.
With more efficient execution of the epigenetic feedback regulation, more attractors with different expression patterns, i.e., more cell types, are generated.  
In some networks, the initial expression levels are simply embedded into epigenetic modifications, whereas for other networks, mutual feedback between expression levels and modifications bring about hierarchically ordered attractors from an oscillatory state.
In such a case, the attractor number shows non-monotonic change against the  the rate of epigenetic feedback regulation $v$.
The mechanism of non-monotonic dependency on $v$, i.e., the attractor generation and pruning, is explained in terms of dynamical systems theory. 

By using the change in expression dynamics under the slow epigenetic modification process, Waddington's epigenetic landscape is explicitly depicted, in which the landscape axis ($X$ axis in Fig.\ref{Waddington}) is given by the principal component of the expression pattern; the depth, $Y$ axis, is given by the developmental time with slow epigenetic modification; and the height is given by $-\log(P(X))$ with $P(X)$ as the cell-number distribution of $X$.
In particular, when the original attractor in the absence of epigenetic modification is a limit cycle, the timing of branching to different cell types, number of differentiated cell types, and number fraction of each cell type are all robust to perturbations during the course of development and to the variation of initial conditions.
Hence, the generated landscape satisfies the three postulates implicitly assumed in Waddington's landscape:
(i) hierarchical branching is supported by the hierarchical attractor generation from the limit cycle; (ii) homeorhesis is supported since this branching process is independent of initial conditions and robust to noise; and (iii)the cell-number robustness is demonstrated since $P(X)$ is also independent of initial conditions and robust to noise.
This robustness in the path and in the cell-number distribution to perturbation is an essential requirement for the development of multicellular organisms \cite{alsing2013differentiation}.

Our theoretical model assumes epigenetic feedback regulation.
Although the transient modification in epigenetic factors has been experimentally confirmed
\cite{kangaspeska2008transient, rulands2018genome}, the extent to which this modification depends on gene expression is not yet clearly elucidated.
Considering that epigenetic change stabilizes the cellular states, it is rather natural to assume positive feedback from the expression level to modification, i.e., if expressed (repressed), it is easier (harder) to be expressed, whereas some molecular mechanisms for such positive feedback have been suggested \cite{dodd2007theoretical, sneppen2008ultrasensitive}.
However, direct evidence as well as quantitative estimates for the time scale of epigenetic change require further experimental elucidation in the future.

The significance of oscillation in the cellular state for the differentiation process was previously discussed \cite{suzuki2011oscillatory}.
Indeed, the cell state is not fixed but rather involves several oscillatory modes, including circadian and cell-division cycles.
Furthermore, oscillatory expression has recently been uncovered for embryonic stem cells \cite{kobayashi2009cyclic, imayoshi2013oscillatory, aulehla2008oscillating, canham2010functional}, which is ultimately lost in cells committed to differentiation.
Note that the relevance of an oscillatory state to pluripotency was previously discussed in the context of an alternative approach to the epigenetic landscape with respect to inclusion of cell-cell interactions \cite{miyamoto2015pluripotency}.
In this case, the initial oscillation in expression levels is lost with an increase in the cell number and resulting amplification of cell-cell interactions accordingly.
Hence, the two approaches, i.e., cell-cell interactions and epigenetic modification, are compatible.
Indeed, a model that includes both approaches was previously investigated, in which epigenetic modification of several genes such as Oct4 and Nanog leads to the commitment of cells from an undifferentiated state, which is consistent with experimental observations \cite{niwa2000quantitative, boyer2005core, dunn2014defining}.

The canalization in Waddington's landscape is valid for the normal developmental process.
However, through certain external operations, the path of committed cells can be reversed to an undifferentiated state in a process known as reprogramming \cite{takahashi2006induction, hochedlinger2009epigenetic, hajkova2008chromatin, huang2009reprogramming}.
In the present model, by externally overexpressing some genes for a given timespan, the threshold $-\theta_i$ that was initially increased can be decreased so that the expression level recovers, which matches the experimental procedure used to create induced pluripotent stem cells.
In the future, it will be important to elucidate the condition required for such reprogramming by identifying the specific genes in the network that need to be overexpressed so as to climb up to the most upstream location in the landscape under the present theoretical framework.

The generation and pruning of attractors that depend on the epigenetic feedback rate is itself an interesting phenomenon in terms of dynamical systems of both fast and slow elements, which requires an analysis beyond the breadth of adiabatic elimination \cite{aoki2013slow}. That is, if the time scales are clearly separated, the change in fast expression would be represented as an attractor change against the slow epigenetic state as a control parameter.
In contrast, mutual feedback between the two is important, as shown in the present study with regard to the interaction between the nullclines and the variables.
Therefore, an appropriate analytical method that is capable of capturing such feedback dynamics needs to be developed.

Homeostasis, robustness of a steady state in biological systems has gathered much attention over decades.
This, for instance, has been discussed as the stability of the final state (attractor) against perturbations.
On the other hand, homeorhesis concerns with the stability of the time course of a state, against the change in the initial conditions or perturbations.
So far,  studies on homeorhesis  are rather limited: Few examples include relaxation dynamics in signal transduction process independent of the initial condition \cite{young2017dynamics}, robust developmental process with cell-cell interaction \cite{kaneko1999isologous, suzuki2011oscillatory}, and robust ecological dynamics in an experiment consisting of algae and ciliates \cite{chuang2019homeorhesis}.
For homeorhesis to work, existence of slower time scale and buffering of initial variation will be needed. The hierarchial attractor generation by slower epigenetic feedback after attraction to a limit cycle will provide one general mechanism for the homeorhesis.

\begin{acknowledgments}
The authors would like to thank Chikara Furusawa and Tetsuhiro Hatakeyama for stimulating discussions.  This research was partially supported by a Grant-in-Aid for Scientific Research (S) (15H05746) from the Japanese Society for the Promotion of Science (JSPS) and a Grant-in-Aid for Scientific Research on Innovative Areas (17H06386) from the Ministry of Education, Culture, Sports, Science and Technology (MEXT) of Japan.
\end{acknowledgments}

\nocite{*}

\bibliography{apssamp}

\end{document}